\documentclass[12pt,a4paper]{article}
\usepackage{amsmath}
\usepackage{cite}
\begin{document}
\title{ Annotations regarding the \textit{Oxford Questions}}

\author{Spiridon  Dumitru\\ 
(Retired)Department of Physics, \textit{"Transilvania"} University,\\
B-dul Eroilor 29, 500036 Bra¡sov, Romania,\\
 E-mail: s.dumitru42@yahoo.com }

\date{\today}

\maketitle
\begin{abstract}
 The recently published "Oxford Questions" are supplemented with annotations concerning: doctrine 
 of wave packets collapse (and subsidiarily   Schrodinger's cat thought experiment), description 
 of quantum measurements respectively interpretation of uncertainty   relations.
\end{abstract}
\section{Introduction}
Highly authorized opinions call attention to the major deficiencies of  modern debates and researches referring to the foundations  of Quantum Mechanics (QM). As such opinions (\textbf{o})  can be quoted the next ones :  

$\textbf{o}_1$ :  \textit{"the idea that there are defects in the foundations of orthodox quantum theory is unquestionable present in the conscience of many physicists"} \cite{1} 
 
$\textbf{o}_2$ : \textit{"There is now . . . no entirely satisfactory interpretation of quantum mechanics"} \cite{2}.\\
That is why it was a remarkable initiative to seek a \textit{"formulation of a list of main open questions about the foundations of quantum physics"} \cite{3}. Such a list was elaborated within a  distinguished conference and its items are known now as \textit{"Oxford Questions"} (OQ) \cite{3}.

Of course, for the future,  it is expected that OQ will motivate more or less extensive actions (debates and adequate researches). But, for a first stage of the before-mentioned actions, can be  of some interest to supplement OQ with few annotations prefigured by investigations already noticed (in publications).   Such a supplementation is the aim of the present short paper. So we try to bring into attention some aspects regarding : (i) Wave Packet Collapse  (WPC) doctrine (and subsidiarily   Schrodinger's cat thought experiment), (ii)   Quantum  Measurements   (QMS)  description  and  (iii) interpretation of Uncertainty Relations (UR). We hope that the respective aspects can be useful for further  studies  stimulated by OQ.              
\section{Ephemeral character (caducity) of the doctrine about wave packet collapse}
       Historically WPC doctrine was brought into the scientific debates by the conflict between the following two suppositions (\textbf{s}) 
       
        $\textbf{s}_1$ : 'the old opinion that a measurement of a QM observable  should be regarded as a single sampling (trial) which gives an unique deterministic value'   
        
        $\textbf{s}_2$ : 'the agreement, enforced by theoretical usage, that for a QM observable regarding a  state of a quantum system  one  resorts to probabilistic (non-deterministic) entities represented by an operator  respectively by a wave function/packet'. 

For avoiding conflict   between suppositions ($\textbf{s}_1$) and ($\textbf{s}_2$) it was invented the conception that, during a QMS, the wave packet/function collapse into particular eigenfunction   associated to a unique  deterministic  value of measured observable.  Such a conception led to  the WPC doctrine regarded as a quasi-dogma. The respective doctrine  was assumed, in different readings, within   a large number of mainstream publications (see \cite{2,4,5,6} and references).  But, as a rule, the before-mentioned assumptions were (and still are) not accompanied with adequate elucidations concerning the truthfulness of the alluded doctrine in relation with the  natural themes of QM.  Now OQ \cite{3}  put forward the matters (\textbf{m}):

$\textbf{m}_1$ :\textit{"whether or not the 'collapse of the wave packet' is a  physical process"} 
 
$\textbf{m}_2$ :\textit{"How can the progressive collapse of the wave function be experimentally
 monitored?" }

$\textbf{m}_3$ : ' by which theoretical  scheme  can be described WPC ?'.
\newpage
 As annotations to the above mentioned OQ matters $\textbf{m}_1$ - $\textbf{m}_3$ now we wish to bring into attention the ideas prefigured and argued in our paper \cite{7}. In the main we pointed out the ephemeral  character   (i.e. caducity) of WPC doctrine. Principally our argumentations are grounded on the following indubitable facts.  Mathematically a quantum observable (described by a corresponding operator) is a true random variable. In a theoretical framework, for a given quantum state/system, such a variable must be regarded as endowed with a spectra of values associated with corresponding probabilities (more exactly with a wave function/packet as probability-amplitude). Then, from an experimental perspective, a measurement of a quantum observable requires an adequate number of samplings finished through a significant statistical group of data/outcomes. That is why one can conclude that the  supposition  ($\textbf{s}_1$) of WPC doctrine appears as a false premise while the whole respective doctrine proves oneself to be an useless speculation. 

       Previously noted   conclusion can be consolidated indirectly by mentioning the quantum-classical probabilistic similarity (see \cite{8,9}) among the QM observables and Macroscopic Random Variables (MRV), studied within thermodynamic theory of fluctuations (e.g. in \cite{10,11,12}).  In its wholeness a MRV is characterized by a continuous spectra of values   associated with an intrinsic   probability density. Then for a MRV a single experimental sampling  delivering  an  unique value (result) is worthlessly. Such a sampling is not described as a collapse of the mentioned probability density. Similarly a QMS must   not  be represented as a WPC. Moreover a true experimental evaluation of a MRV requires an adequate lot of samplings finished through a statistical set of individual results.  A plausible theoretical description of the alluded evaluation   can be done \cite{9,13,14} through an information transmission process. 
       
      Subsidiarily to the above considerations about WPC doctrine can be brought into question some remarks \cite{7} concerning the famous Schrodinger's cat thought experiment.  The essential element in the respective experiment is represented by a killing single decay of a radioactive atom. But the individual lifetime of a single decaying atom is a random variable. That is why the mentioned killing decay is in fact a twin analogue of the single sampling considered in supposition($\textbf{s}_1$)  of WPC doctrine. So, the previous considerations reveal the notifiable fact that is  useless (even forbidden) to design experiments or actions that relies solely on a single deterministic sampling of individual lifetime random variable. Accordingly the Schrodinger's experiment signifies   nothing but just a fiction (figment) without any scientific value. Such a  significance can be consolidated  by observation \cite{7} that it is possible to imagine a macroscopic thought experiment completely analogous  with Shrodinger's one. Within the respective analogue a classical (macroscopic) cousin of the Schrodinger's cat can be killed through a  launching of a single macroscopic ballistic projectile. More specifically the killing macroscopic device is activated by the reaching of the projectile in a probable hitting point. But the respective point of   has  characteristics of a true random variable. Consequently   the launching of a single projectile is a false premise similar
 with the supposition  ($\textbf{s}_1$) of WPC doctrine. Add here the known fact that in practice of the traditional artillery (operating only with ballistic projectiles but not with propelled missiles) for destroying a military objective one uses a considerable (statistical) number of projectiles but not a single one. So the whole situation of the non-quantum cousin is completely analogous with the one of quantum Schrodinger's cat. Therefore the thought experiment with classical cousin makes evident oneself as another fiction without any real significance. 
 
  Taking into account the above mentioned indubitable arguments and facts we think that in a natural understanding   the "collapse   of the wave packet"   can't      be considered   as a physical process.  Therefore the further debates and researches expected to be roused by OQ   would be most appropriate to ignore the elements regarding   WPC doctrine. Particularly it must be regarded as being worthlessly  allegations such is : \textit{"the Schrodinger's cat thought experiment remains a topical touchstone
for all interpretations of quantum mechanics"}. Note that similar allegations are present in many science popularization texts, e.g. in the ones disseminated via the Internet. 
\section{	Description of quantum measurements}
       As an another annotation to the OQ   can be taken into account some remarks about QMS in relation with   QM theory as they were suggested and pointed out in \cite{7,9,14}. Firstly it should be noted the fact   that, physically, a QMS must be regarded according to the observations from the previous section. That is why description of QMS have to be approached as an scientific branch separate and additional to the usual version of QM. This   because, on the one hand, usual QM deals exclusively only with the representation   of intrinsic properties of   the studied systems. On the other hand QMS must to contain obligatorily some theoretical (mathematical)   models about practical characteristics of measuring devices/procedures. Choices of the mentioned models may depend on the alluded characteristics as well on   some mathematical considerations. In \cite{7,9,14}, similarly with the description \cite{13} of measurements for MRV,  we suggested that a QMS have to be viewed  as an information transmission process. In such a process the measured system appears as information source while the measuring device plays the role of a information transmission channel.  Part of the mentioned process the QM operators (describing quantum observables) preserve their mathematical expressions. Additionally   the transmission of   quantum probabilistic attributes is described by means of linear transformations for probability density and current
(associated with the corresponding wave function/packet).

\section{Few words about uncertainty relations}
   In our days   one can say without exaggeration that \cite{15}   \textit{"uncertainty 
   principle  ...  epitomizes quantum physics, even in the eyes of the scientifically informed public"},  the  respective principle being symbolized by the famous Uncertainty Relations (UR). Contiguously in \cite{3}  OQ are associated with the remark that as regards QM  \textit{"foundational studies since the 1960s, any list of highlights must surely include ... various deeper analyses, ..., of uncertainty relations"}.  That is why a succinct    survey of UR   topic   can offer nontrivial elements for an annotation to the  OQ suggestions. 
   
 Here, inspired by the ideas promoted in  papers \cite{9,16}, we try to present a survey on the mentioned kind. The story of  UR  subject  started from the wish to adopt an unique and generic interpretation for the Heisenberg's thought-experimental  formula  respectively for the theoretical relation due to Robertson and  Schrodinger.  So it was  ratified  the so called  Conventional Interpretation of UR (CIUR) which, in the main, can be expressed \cite{9,16} through  few basic items. But \cite{9,16}, the respective items  proves themselves to be nothing but wrong sentences. More precisely : (i) Firstly, the Heisenberg's thought-experimental  formulas  remain as provisional fictions destitute of durable physical significance, (ii) Secondly, the Robertson - Schrodinger relations are simple mathematical interconnection regarding QM fluctuations, from the same family with the formulas concerning fluctuations of  MRV and (iii) Thirdly the existence \cite{8,9,10,11}of a quantum-classical probabilistic similarity thanks to which \cite{8,9} the Planck constant $\hbar$ has an authentic classical analog represented by the Boltzmann constant $k_B$ , both  $\hbar$ and $k_B$  being similar generic indicators of stochasticity (randomness). Then CIUR is deprived of necessary qualities of a valid scientific construction and must be abandoned as a wrong conception without any real value or scientific significance.  So one finds a  class of solid arguments which come to advocate and consolidate the Dirac's intuitional guess \cite{17}  that: \textit{"uncertainty relations in their present form will not survive in the physics of future"}.
 
The above presented survey about UR has potentiality   to draw attention to certain elements which can be incorporated into studies that will be stimulated by OQ.
\section{	Closing thoughts }
We appreciate that, for the studies (debates and researches) which will germinate and grow due to OQ,  can be of some nontrivial interest the ideas outlined in above annotations and prefigured in  our papers \cite{7,8,9,10,11,12,13,14,16} (certain of them being freely accessible from the mentioned sites). So our hope is that such  previsioned studies will be evaluated   the real value and utility of the mentioned ideas.

\end{document}